\documentclass[twocolumn,showpacs,preprintnumbers,amsmath,amssymb,prb]{revtex4}

\usepackage{graphicx}
\usepackage{bm}
\usepackage{subfigure}

\begin{document}

\title{Detecting Majorana fermions by nonlocal entanglement between quantum dots}
\author{Zhi Wang and Xiao Hu}
\address{WPI Center for Materials Nanoarchitectonics, National Institute for Materials Science,
Tsukuba 305-0044, Japan}

\date{\today}

\begin{abstract}
Nonlocal entanglement between two quantum dots can be generated
through Majorana fermions. The two Majorana fermions at the ends of
an one-dimensional topological superconductor form a nonlocal fermion
level, coupling to the occupation states of two quantum dots put
close to the two ends, and the entire system will come into an
entangled state.  After introducing a charging energy by a
capacitor, entanglement of the entire system can manifest itself
through the nonlocal entanglement between the two quantum dots. That
is, when measuring the electron occupations of the quantum dots, the
measurement result of one quantum dot will influence the measurement
result of the other quantum dot. This nonlocal entanglement between
the two quantum dots is a strong evidence of the nonlocal nature of
the fermion level constructed by two Majorana fermions.
\end{abstract}


\maketitle

\noindent {\it Introduction.--} The existence of Majorana fermions (MFs) in solid
state systems has been under intense investigation recently
\cite{kitaev,read,ivanov,fu,sarma,sau,lutchyn,oreg,lutchyn2}. It attracts so much
attention mainly because they might be useful for quantum computation\cite{kitaev2}.
 Two well separated MFs define a nonlocal fermion level, which can be either occupied or
 empty, and thus form a topological qubit storing quantum information nonlocally.
 This nonlocal topological qubit may well be incorporated with the standard
 superconducting qubit, and used for decoherence-free quantum computing\cite{jiang,bonderson}.

One of the candidates which support MFs is the spinless p-wave
superconductor. In these topological superconductors, MFs may arise
due to the particle-hole redundancy\cite{kanermp}. MFs have already
been predicted in several superconducting proximity systems which
resemble the spinless p-wave
superconductor\cite{kitaev,read,ivanov}, including
Sr$_2$RuO$_4$\cite{sarma}, interface between an s-wave
superconductor and a topological insulator\cite{fu}, and spin-orbit
coupling nanowire in proximity to an s-wave
superconductor\cite{sau,lutchyn,oreg}.

Two obvious issues concerning MFs are how to detect and manipulate
them. A number of proposals have been suggested to detect MFs in
superconducting systems, mostly by tunneling
experiments\cite{lee,fu2,lutchyn,akhmerov,Ioselevich,alicea,tsvelik}. One
unique property of the MFs is that two of them can form a nonlocal
fermion level. Making use of this property,  it has been proposed by
Fu\cite{fu3} that there can be a novel electron teleportation
across the superconductor when it is connected to the
ground through a capacitor, as an evidence of MFs existence.
Meanwhile, a quantum dot (QD) close to the end of a one dimensional topological superconductor was introduced by Flensberg to manipulate the
MFs\cite{flensberg}. In his setup, an electron can tunnel from the
QD into the superconductor and generate a ground state rotation,
which is similar to braiding the two MFs. It was also illustrated
that quantum information may be transferred between a topological
qubit formed by MFs and a spin qubit formed by QD\cite{flensberg2}.

In this work, we adopt the system schematically shown in Fig.~1,
where two QDs are coupled by tunneling to two well separated MFs
supported by a one-dimensional spin-orbit coupling nanowire. The
nanowire is in proximity to a mesoscopic superconductor which is
connected to the ground through a capacitor. When the energy levels
of the two QDs are tuned close to the fermion level formed by MFs,
the entire system will come into an entangled state. We
show explicitly that, because of the charging energy, nonlocal
entanglement between the two QDs appears, i.e., the measurement
result of electron occupation on one QD will influence the
possibility of finding one electron on the other QD. Therefore,
detecting the nonlocal entanglement of the two QDs  will provide an
unambiguous evidence of the nonlocal fermion level, and thus prove
the existence of MFs.

\begin{figure}
\begin{center}
\leavevmode
\includegraphics[clip=true,width=0.9\columnwidth]{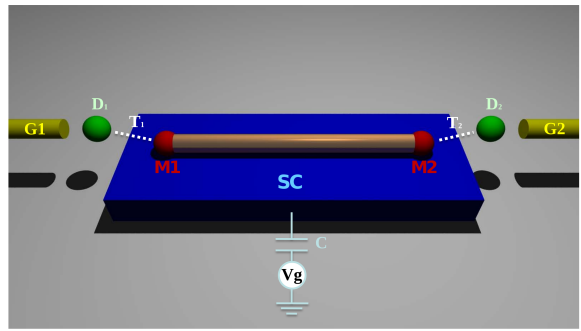}
\caption{Tunneling setup of two quantum dots and a one-dimensional
spin-orbit coupling nanowire with two Majorana fermions at its
ends.} \label{fig:1}
\end{center}
\end{figure}

\vspace{3mm} \noindent {\it Model.--} The system we consider is a
one-dimensional  spin-orbit coupling wire in proximity with an
s-wave superconductor. Under proper magnetic field, the system resembles
the properties of spin-less p-wave topological
superconductivity\cite{sau,oreg}. At the two ends of the wire, there
are zero energy states described by the Bogoliubov-de-Gennes
equation. These zero energy modes has been identified theoretically
as local Majorana bound states\cite{sau,oreg}. The two Majorana
operators $\gamma_L$ and $\gamma_R$ are defined
as\cite{oreg,fu3,flensberg}

 \begin{eqnarray}
\gamma_{L,R} \equiv \int dx \bigg(  e^{-i\phi/2}[ \xi_{L,R}(x) c^{\dagger}_{\uparrow}(x)+\eta_{L,R}(x) c^{\dagger}_{\downarrow}(x)]  \nonumber\\\ +e^{i\phi/2} [\xi^*_{L,R} (x)c_{\uparrow}(x)+\eta^*_{L,R}(x) c_{\downarrow}(x)] \bigg),
\end{eqnarray}
with $\phi$ the superconducting phase, $\xi$ and $\eta$ bound state wave functions centered at ends of the wire, $c^{\dagger}$ and $c$ electron creation and annihilation operators. We assume that the distance between these two Majorana bound states is large enough, as such we can ignore their interaction. The existence of these two Majorana bound states indicates that the superconducting ground state of the topological superconductor may have either even or odd number of electrons\cite{fu3}.

Now we introduce two quantum dots, one close to each end of the
wire. The QDs are coupled to the two MFs through tunneling. We set
the QD in the Coulomb blockade regime, i.e., its level spacing is
large enough so that it may be modeled as one energy level. Since a
large magnetic field is applied in order to achieve the MFs, it is
enough to consider one spin direction for the electrons on QDs.  In
this case, the QDs can be described by a spin-less electron energy
level Hamiltonian\cite{flensberg},

\begin{eqnarray}
H_{QD}=\sum_{j=L,R}\epsilon_{j} c^{\dagger}_{j}c_{j}.\end{eqnarray}

The QDs are coupled to the wire through electron tunneling between QDs and Majorana bound states, with the tunneling Hamiltonian \cite{fu3},

 \begin{eqnarray}
H_T=\sum_{j=L,R}  T_j c^{\dagger}_j \gamma_j e^{-i\phi/2} +h.c.
\end{eqnarray}
Here the Majorana operator $\gamma_{L,R}$ changes the fermionic parity of the superconductor\cite{fu3}.

The superconductor is considered of mesoscopic size, and connected
to the ground by a capacitor. Then the system will have a charging energy term\cite{fu3},

 \begin{eqnarray}
H_C= (ne-Q_0)^2/2C.
\end{eqnarray}

In the following, we consider the superconducting energy gap to be the
largest energy scale and set it as infinity, thus the superconductor is in its ground state and the superconducting quasiparticles are irrelevant, which is a good approximation as far as low temperature is concerned. Now we have the total Hamiltonian of the system,
 \begin{eqnarray}
H=H_{QD}+H_C+H_T.
\end{eqnarray}
 For simplicity, we set the same energy level for the two QDs $\epsilon_L=\epsilon_R=\epsilon$, as well as the tunneling strength $T_L=iT_R=T$ .

We initialize the system with $\epsilon=-\infty$, with both QDs
occupied by one electron as denoted by $|1 \rangle_L$ and
$|1\rangle_R$. Without losing generality, we set $Q_0= 2N_0
e$, which enforces the superconductor accommodate $2N_0$
electrons at the ground state denoted\cite{fu3} by $|2N_0\rangle_S$.
Since the Hilbert space of the whole system is simply the direct
product of electron number states of QDs and the superconductor, our initial ground state is $|1\rangle_L
|2N_0\rangle_S |1\rangle_R$.

Now we adiabatically increase the QD energy level $\epsilon$. When
$\epsilon$ approaches zero, tunneling between QDs and MFs becomes
important. Because the QD levels are fully occupied, electron
tunneling to the QDs is forbidden. On the other hand, the electron
may tunnel from either one of or both QDs to the
superconductor via MFs, leading to four possible states: $
|1\rangle_L|2N_0\rangle_S|1\rangle_R$ for no electron tunneling,
$|1\rangle_L|2N_0+1\rangle_S|0\rangle_R \equiv -i e^{i \phi/2}
\gamma_{R} c_{R}   |1\rangle_L|2N_0\rangle_S|1\rangle_R$ for the
electron on the right QD tunneling to the superconductor,
$|0\rangle_L|2N_0+1\rangle_S|1\rangle_R \equiv  e^{i \phi/2}
\gamma_{L} c_{L}    |1\rangle_L|2N_0\rangle_S|1\rangle_R$ for the
electron on the left QD tunneling to the superconductor, and finally
$|0\rangle_L|2N_0+2\rangle_S|0\rangle_R \equiv -i e^{i \phi/2}
\gamma_{R} c_{R}  e^{i \phi/2} \gamma_{L} c_{L}
|1\rangle_L|2N_0\rangle_S|1\rangle_R $ for the two electrons on two
QDs tunneling to the superconductor. We notice that since only one
electron originally exists on each QD, there is no other possible
states due to the charge conservation.
  When $\epsilon$ is adiabatically increased, the ground state will gradually evolve away from the initial state $|1\rangle_L|2N_0\rangle_S|1\rangle_R$ due to the electron tunneling,
and achieve a superposition state of the four different electron occupations. Nonlocal entanglement between the two QDs can be achieved as revealed below.

\begin{figure}
\includegraphics[scale=0.4]{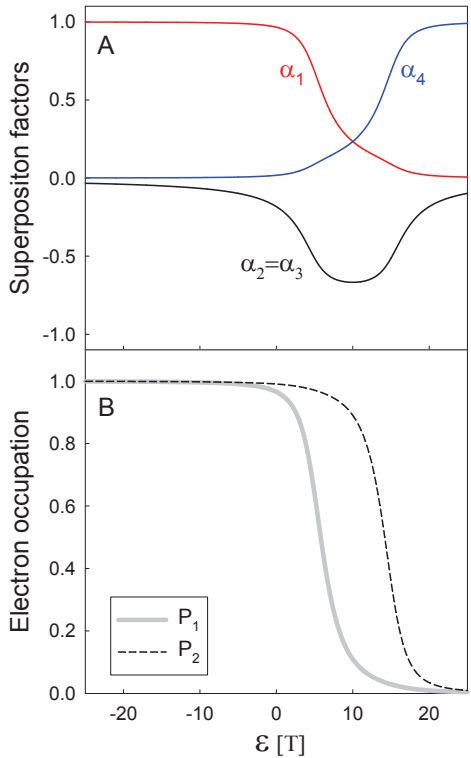}
\caption{Dependence of superposition factors (A) and electron
occupation on the right QD (B) on the energy level of the QDs
$\epsilon$, for $\frac{e^2}{2C}=5T$. $P_1$ and $P_2$ are defined in the main text.}
\end{figure}

Using the four possible states as the basis,

\begin{eqnarray}
|1\rangle_L |2N_0\rangle_S |1\rangle_R  =\left( \begin{array}{c}  1 \\ 0 \\0 \\0 \end{array}\right)
,
|1\rangle_L |2N_0+1\rangle_S|0\rangle_R  =\left( \begin{array}{c}  0 \\ 1 \\0 \\0 \end{array}\right),
\end{eqnarray}

  \begin{eqnarray}
|0\rangle_L |2N_0+1\rangle_S |1\rangle_R  =\left( \begin{array}{c}  0 \\ 0 \\1 \\0 \end{array}\right)
,
|0\rangle_L |2N_0+2\rangle_S |0\rangle_R  =\left( \begin{array}{c}  0 \\ 0 \\0 \\1 \end{array}\right),\nonumber
\end{eqnarray}
the Hamiltonian (5) is reduced to a 4 by 4 matrix. The matrix elements can be given straightforwardly based on the
definitions of states, for example,

  \begin{eqnarray}
H_{24}   = && \langle 0 |_R \langle 2N_0+2|_S \langle 0|_L \hat{H} |1\rangle_L|2N_0+1\rangle_S |0\rangle_R \nonumber \\\
  =  &&T  \langle 1 |_R \langle 2N_0|_S \langle 1|_L  c^{\dagger}_{L} \gamma_{L} e^{-i \phi/2}
  c^{\dagger}_{R} \gamma_{R} e^{-i \phi/2}    \nonumber\\\ &&
   e^{i \phi/2} \gamma_{L} c_{L} e^{i \phi/2} \gamma_{R} c_{R}
   |1\rangle_L|2N_0\rangle_S |1\rangle_R =T.
\end{eqnarray}
The 4 by 4 matrix is thus given by,

  \begin{eqnarray}
H  =\left( \begin{array}{cccc}
  2\epsilon  &T &T&0
 \\T^*&\epsilon+\frac{e^2}{2C}&0&T
 \\T^*&0&\epsilon+\frac{e^2}{2C}&T
 \\0 &T^*&T^*&\frac{2e^2}{C}
 \end{array}\right).
\end{eqnarray}
Here we note that the same matrix can be obtained for the
case $Q_0= (2N_0+1) e$ for the  corresponding basis.

The lowest eigenvalue of this matrix is the ground state energy, and its corresponding eigenvector is the ground state.  It is clear that when $\epsilon=-\infty$, the ground state is  $|1\rangle_L|2N_0\rangle_S|1\rangle_R$. Now we calculate the ground state with $\epsilon$ adiabatically increased.  The problem is to diagonalize the matrix and find the ground state. In general, this ground state is a superposition state of the four possible states,
  \begin{eqnarray}
|G\rangle=\alpha_1|1\rangle_L |2N_0\rangle_S |1\rangle_R+\alpha_2|1\rangle_L |2N_0+1\rangle_S |0\rangle_R\nonumber\\\ +\alpha_3|0\rangle_L |2N_0+1\rangle_S |1\rangle_R+\alpha_4|0\rangle_L |2N_0+2\rangle_S |0\rangle_R,
\end{eqnarray}
where the superposition factor
$(\alpha_1,\alpha_2,\alpha_3,\alpha_4)$ is the eigenvector of the
matrix associated with the lowest eigenvalue.

 \begin{figure}[t]
\includegraphics[scale=0.4]{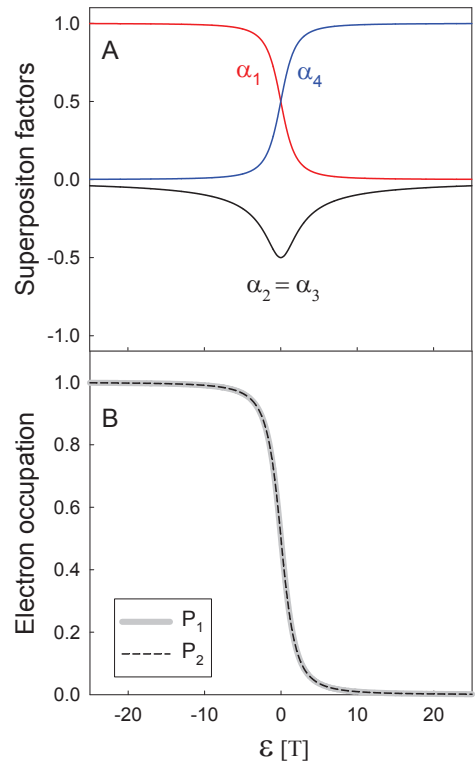}
\caption{Same as Fig. 2 except for $\frac{e^2}{2C}=0$.}
\end{figure}

\vspace{3mm} \noindent {\it QD entanglement.--} The
eigenvalues and eigenvectors of a 4 by 4 matrix can be found
analytically with lengthy expressions except some special cases.
Here we choose to present the result numerically for simplicity.
Using T as the unit and setting $\frac{e^2}{2C}=5T$, we show the
four components of the eigenvector of the lowest eigenvalue evolving
with $\epsilon$ in Fig. 2A. It is clear that all factors but
$\alpha_1$ vanish when $\epsilon\rightarrow -\infty$, whereas all
factors but $\alpha_4$ vanish when $\epsilon \rightarrow \infty$. In
these two limit cases, there is no entanglement in this system.
However, when $\epsilon$ approaches zero energy, all factors are
non-zero, thus entanglement appears in the system.

Next, we show explicitly that the entanglement of the entire system
can be exposed and detected by nonlocal entanglement between the two
QDs. For this purpose, we calculate the probability of finding one
electron on the right QD with electron occupation on the left QD
$P_1= {\alpha_1^2}/({\alpha_1^2+\alpha_2^2})$, and that
 without electron occupation on the left QD $P_2= {\alpha_3^2}/({\alpha_3^2+\alpha_4^2})$,
as shown in Fig. 2B.
 When $\epsilon$ is positively or negatively large, the two curves overlap, which means the measurement result on the left QD has no influence on the measurement result on the right QD. However, when $\epsilon$ approaches zero, the two curves deviate from each other, implying that the measurement result of the left QD will influence the measurement result of the right QD. This is the signature for entanglement between these two QDs.

Now, let us discuss the importance of the capacitor. For this
purpose, we investigate the case of infinite capacitance, with the
results of superposition factors corresponding to the ground state
presented in Fig. 3A. At the first glance, the results for
$C=\infty$ seem similar to those in Fig. 2, namely, only $\alpha_1$
or $\alpha_4$ is non-vanishing when $\epsilon$ is negatively or
positively large; all factors are non-vanishing around zero energy,
implying entangled ground state for the entire system.

 \begin{figure}
\includegraphics[scale=0.4]{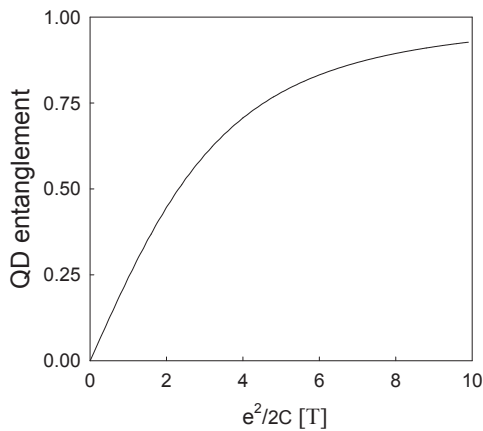}
\caption{Capacitance dependence of entanglement between the
two QDs measured by the maximum value of $P_2-P_1$.}
\end{figure}

However, although the entire system will come into an entangled
state around $\epsilon=0$, there is no {\textit{direct}}
entanglement between left and right QDs in the present case of
$C=\infty$. In order to clarify this point, we calculate the
possibility of finding one electron in the right QD, with and
without the electron occupation in the left QD. As seen in Fig. 3B,
the two curves overlap totally, even when $\epsilon$ approaches
zero. It demonstrates that there is no measurable nonlocal
entanglement induced between the two QDs, although the entire system
is in an entangled state.

The key point here is that, for $C=\infty$ or equivalent a bulk
superconductor, one has to detect the fermionic parity of the
superconductor to reveal the existence of the Majorana
fermions, which is still a difficult task. By introducing the
capacitor to a mesoscopic superconductor, the nonlocally entangled
MFs can be exposed by simply detecting the
nonlocal entanglement between the QDs.

As a measure for the
entanglement between the two QDs, we calculate the maximal
difference between the measurement results $P_2-P_1$ as a function
of the capacitance. As shown in Fig. 4, the degree of entanglement
increases with decreasing capacitance monotonically, with $e^2/2C$
not exceeding the superconducting energy gap.

Lastly we discuss about possible experimental implementations of our
idea. Our method does not require any measurement on the
state of MFs and the topological superconductor. All
we need is the measurement of the electron occupation on QDs. Charge
sensing on QD is a well developed area\cite{hanson}, and the
measurement of charge occupation on QD has already been achieved
with integrated radio-frequency single-electron transistor\cite{lu}.

\vspace{3mm} \noindent {\it Summary.--} In summary, we
have investigated the tunneling system of two quantum
 dots and a one-dimensional topological superconductor with two Majorana fermions
 at the two ends. When the quantum dot energy level is close to zero
 energy the system exhibits an entangled state.  When the superconductor
 of mesoscopic size is connected to  the ground through a capacitor,
 the entanglement of the whole system will show up through the entanglement
 between the two quantum dots.
 Namely, the measurement result of occupation on one quantum dot will
 influence the measurement result on the other quantum dot. This nonlocal
 entanglement between the two quantum dots is a strong evidence of the
 nonlocal nature of the fermion level constructed by the two Majorana
 fermions.

This work was supported by WPI Initiative on Materials
Nanoarchitectonics, MEXT, Japan, and partially by Grants-in-Aid for
Scientific Research (No.22540377), JSPS, and CREST, JST.

\end{document}